\def\dev{\partial}
\def\G{\Gamma}
\def\m{\mu}
\def\n{\nu}

\def\l{\lambda}
\def\T{\Theta}

\def\pmb#1{\setbox0=\hbox{#1}%
  \kern-.025em\copy0\kern-\wd0
  \kern.05em\copy0\kern-\wd0
  \kern-.025em\raise.0433em\box0}

\def\bfxi{\pmb{$\xi$}}
\documentstyle[aps,preprint]{revtex}
\topmargin = - 0.9 cm
\begin{document}
\begin{titlepage}
\begin{flushright}
MIT-CTP\#2400\\
IFUP-TH-1/95
\end{flushright}
\vskip 1truecm
\begin{center}
\Large\bf
Fermi-Walker gauge in 2+1 dimensional gravity.
\footnote{This work is  supported in part by D.O.E. cooperative
agreement DE-FC02-94ER40818 and by M.U.R.S.T.}
\end{center}
\medskip
\begin{center}
Pietro Menotti\\
{\small\it
Dipartimento di Fisica dell' Universit\'a, Pisa 56100, Italy and\\[-5.5pt]
I.N.F.N., Sezione di Pisa.}\\[-4pt]
{\small \tt e-mail: menotti@ibmth.difi.unipi.it}\\
{\small\rm  and}\\
Domenico Seminara\\
{\small\it
Center~for~Theoretical~Physics,~{Laboratory~for~Nuclear~Science~and~
Department~of~Physics}\\[-4pt]
Massachusetts Institute of Technology, Cambridge, MA 02139 U.S.A.,}\\[-4pt]
\small\it and  I.N.F.N., Sezione di Pisa.\\ [-4pt]
{\small \tt e-mail: seminara@pierre.mit.edu}
\end{center}
\begin{center}
January 1995
\end{center}

\end{titlepage}

\begin{abstract}
It is shown that the Fermi-Walker gauge allows the general solution of
determining the metric given the sources, in terms of simple quadratures.
We treat the general
stationary problem providing explicit solving formulas for the metric
and  explicit support conditions for the energy momentum tensor.
The same type of solution is obtained for the time dependent problem
with circular symmetry. In both cases the solutions are classified in
terms of the invariants of the Wilson loops outside the sources.
The Fermi-Walker gauge, due to its physical nature, allows to exploit the
weak energy condition and in this connection it is
proved that, both for open and closed universes with rotational
invariance, the energy condition  imply the
total absence of closed time like curves.

The extension of this theorem to the general stationary problem, in
absence of rotational symmetry is considered. At present such extension is
subject to some assumptions on the behavior of the determinant of the
dreibein in this gauge.

\end{abstract}
\section{Introduction.}
Gravity in 2+1 dimensions \cite{DJH} has provided a good theoretical
laboratory
both at the classical and quantum level. On the classical side most
attention has been devoted up to now to stationary solutions in
presence of point-like or string-like sources \cite{DJH} \cite{grilee}.
Many interesting
features have emerged. In particular the discovery by Gott \cite{gott}
of systems
of point spinless particles in which closed time-like curves (CTC) are
present has revived the problematics of causal consistency in general
relativity.

In ref.\cite{MS1,MS1b,MS2,MS3} it was shown that the adoption
of gauges of a radial
type \cite{modtol} allows to write down the general
solution of Einstein equations
by means of simple quadratures in term of the sources, given by the
energy momentum tensor. On the other hand gauges of pure radial
nature, even though useful to discuss certain special problems, present
the disadvantage of singling out a special event in space time. So
the Fermi-Walker \cite{ferwal} gauge while retaining all interesting
features of the radial gauges shows the advantage to be applicable both
to stationary and time dependent situations. It was also shown in
ref.\cite{MS1b,MS2} that if the problem is stationary (presence
of a time-like Killing vector) or if it possesses axial symmetry
(presence of a Killing vector with closed integral curves that are
space like at space infinity) it is possible to give a complete
discussion of the
support property of the energy momentum tensor. In addition the
physical nature of the Fermi-Walker gauge allows to exploit the energy
condition; this possibility turns out very important to
discuss the occurrence of closed time-like curves \cite{MS2,CFGO}.
With regard to the problem of causality in ref.\cite{CFGO} it was
shown that Gott like
configurations cannot occur in an open universe of spinless point
particles with total time-like momentum; this is an important general
result even though it does
not exclude the occurrence of CTC's.~ 't Hooft
\cite{tH1,tH2} for a system of point
spinless particles in a closed universe gave a construction of a
complete set of Cauchy
surfaces proving that in the evolution of such Cauchy surfaces no CTC
can occur.
Along the lines of the present paper a  result was obtained in
ref.\cite{MS2} for open stationary universes with axial symmetry,
proving that absence of CTC's at space infinity prevents, when
combined with the weak energy condition the occurrence of CTC's
anywhere.

Obviously given the simpler setting of 2+1 dimensional
gravity compared to $3+1$ gravity one should like to have
general simple statements regarding the
problem of CTC's for any universe whose matter satisfy either the weak
or the dominant energy condition but up to now such a general
statement is missing.

The present paper is organized as follows: in Sect.II we write down
 the defining equations of the Fermi-Walker gauge in the first order
formalism that  will be adopted in the sequel of the paper. In
Sect.III we solve the conservation and symmetry
equations for the  energy momentum tensor in the Fermi-Walker
gauge. In Sect.IV we turn to the time dependent problem with axial
symmetry. We solve explicitly the symmetry constraints and write down
the support equations for the energy-momentum tensor in terms of the
Lorentz and Poincar\'e holonomies and show how such conditions can be
explicitly implemented. For the time dependent problem we have no
assurance that the Fermi-Walker coordinate system gives a complete
description of space-time. On the other hand in the general stationary
situation, that is dealt with in Sect.V the completeness of the
Geroch projection assures the completeness (or even overcompleteness)
of the Fermi-Walker coordinate system. That provides a good working
ground for the stationary problem and explicit quadrature formulas are
given and the invariants of the metric are worked out in terms of Lorentz
and Poincar\'e holonomies.
In Sect. VI we turn to the problem of CTC's in the stationary
case. The absence of CTC's is proved for universes which are conical at
space infinity under the hypothesis that the WEC holds and that the
determinant of the dreibein $\det(e)$ in our gauge never vanishes.
The non vanishing of $\det(e)$ can be proved in case of axial
symmetry in the sense that the vanishing of $\det(e)$ implies
either the compactification of the whole space-time or the space
closure of the universe. In this context we are able to extend the
proof of the absence of CTC's for open universes with axial symmetry
also to closed universes with axial symmetry. On the other hand in
absence of axial symmetry, up to now we have no way to dispose of the
hypothesis $\det(e)\neq 0$.

\section{The Fermi-Walker gauge}
The Fermi-Walker (FW)\cite{ferwal}
gauge is considered the natural system of coordinates
for an accelerated observer: it connects in a simple way
 physical observables like
acceleration and rotation to geometrical invariant objects like geodesic
distances. These coordinates are usually defined in terms of their explicit
geometrical construction and the fields are given by a perturbative expansion
around the observer' s worldline\cite{Synge,MTW}.
This feature makes it difficult to handle with
them in practical computations.\\
In this section we follow a different approach to the FW gauge, which allows
us  to recover all their well-known properties and to point out some new ones.
Our starting point is the first order formalism.
Here FW coordinates are defined by \cite{MS1}
\begin{equation}
\label{form1}
\sum_i\xi^i\Gamma^a_{bi}=0
\end{equation}
and
\begin{equation}
\label{form2}
\sum_i\xi^i e^a_i=\sum_i\xi^i\delta^a_i,
\end{equation}
where the sums run only over space indices. (In the following the indices
$i,\ j,\ k,\ l \ {\rm and} \ m$ run over space indices, the quantities in a
generic gauge are denoted by a hat and the quantities without hat are
computed in the FW gauge).\\

First of all we shall discuss the possibility of recovering the usual approach
\cite{Synge,MTW}
as a consequence of  the conditions (\ref{form1}) and (\ref{form2})
by exploiting the geometrical content of them.
We know that the Christoffel symbol are given by
\begin{equation}
\label{Cris1}
\Gamma^\lambda_{\mu\nu}=e^\lambda_a \Gamma^a_{~b\nu}e^b_\mu+
e^\lambda_a\partial_\nu e^a_\mu.
\end{equation}
 From eqs. (\ref{form1}) and (\ref{form2}) we obtain
\begin{equation}
\label{xixi}
\xi^i\xi^j \Gamma^\lambda_{ij}(\xi)=\xi^i \xi^j e^\lambda_a\partial_i e^a_j=
e^\lambda_a(\xi^i \partial_i (e^a_j \xi^j)- e^a_j \xi^j)=0.
\end{equation}
Let us call now $x^\mu(\xi)$ the transformation of coordinates that connects
a generic system $\{x^\mu\}$ to the FW one $\{\xi^\mu\}$; the connections
 are related by the following equation
\begin{equation}
\Gamma^\lambda_{\mu\nu}(\xi)=
\frac{\partial \xi^\lambda}{\partial x^\alpha}\left (
\hat \Gamma^\alpha_{\rho\sigma}(x(\xi))~
 \frac{\partial x^\rho}{\partial \xi^\mu}
\frac{\partial x^\sigma}{\partial \xi^\nu} +\frac{\partial^2 x^\alpha}
{\partial \xi^\mu\partial \xi^\nu}\right ).
\end{equation}
Using equation (\ref{xixi}) we obtain
\begin{equation}
\label{x(xi)}
\xi^i\xi^j\frac{\partial^2 x^\alpha}{\partial \xi^i \partial \xi^j}+
\hat \Gamma^\alpha_{\rho\sigma}(x(\xi))~
 \xi^i \frac{\partial x^\rho}{\partial \xi^i}
 \xi^j\frac{\partial x^\sigma}{\partial \xi^j}=0.
\end{equation}
 From eq. (\ref{form2}) we have \footnote{The information given by eq.
(\ref{Eqq}) is partially contained
in eq. (\ref{x(xi)}). In fact  eq. (\ref{x(xi)}) implies that
$\xi^j\frac{\partial x^\mu}{\partial \xi^j} \hat g_{\mu\nu}(x(\xi))
\xi^i\frac{\partial x^\nu}{\partial \xi^i}$ is an homogeneous function of
degree $2$ in the variables $\xi^i$.}
\begin{equation}
\label{Eqq}
\xi^j\frac{\partial x^\mu}{\partial \xi^j} \hat g_{\mu\nu}(x(\xi))
\xi^i\frac{\partial x^\nu}{\partial \xi^i}=\xi^i g_{ij} (\xi)\xi^j=
\xi^i e^a_i(\xi) \eta_{ab} e^b_j (\xi)\xi^j=-\sum_i \xi^i\xi^i.
\end{equation}
These  two differential equations  define $x^\mu(\xi)$. The exact meaning of
this statement is discussed in appendix A. Obviously it is
impossible to write down the general solution, but the most interesting
properties can be derived without solving eqs. (\ref{x(xi)}) and (\ref{Eqq}).
In fact let us define $x^\mu (\lambda)$ by
\begin{equation}
x^\mu (\lambda)\equiv x^\mu(\xi^0,\lambda \xi^i)
\end{equation}
where $\xi^0$ and $\xi^i$
are kept constant, then from eq. (\ref{x(xi)}) we
obtain
\begin{equation}
\frac{d^2 x^\alpha}{d\lambda^2}+\hat \Gamma^\alpha_{\rho\sigma}(x)
\frac{d x^\rho }{d\lambda}\frac{d x^\sigma }{d\lambda}=0,
\end{equation}
that is $x^\alpha(\lambda)$ are geodesics for each value of
$\xi^0$ and $\xi^i$ and they all start from the curve $s^\alpha(\xi^0)=
x^\alpha(\xi^0,{\bf 0})$. In addition we can also show that these geodesics
are orthogonal to the line $s^\alpha(\xi^0)$.  This result, as we shall see,
is a trivial consequence of the following property of the FW $n$-bein.

For regular fields
(i.e. continuous with bounded derivatives),
\begin{equation}
e^a_i (\xi^0,{\bf 0})=\delta^a_i,\ \ \ e^a_0 (\xi^0,{\bf 0})=\phi(\xi^0)
\delta^a_0\ \ \ {\rm and}\ \ \ \Gamma^a_{~bi} (\xi^0,{\bf 0})= 0,
\end{equation}
if we assume that our $n$-bein is orthonormal, i.e.
$(e^a,e^b)=\eta^{ab}$.\footnote{The presence of this arbitrary function
$\phi(\xi^0)$ is related to the residual gauge invariance. In particular
we can set $\phi(\xi^0)=1$ if
we choose  to parametrize the observer' s world-line with its proper
time.}

In fact taking the derivative of
(\ref{form2}) with respect to $\xi^j$, we have
\begin{equation}
e^a_j (\xi)=\delta^a_j+\xi^i\partial_j e^a_i(\xi).
\end{equation}
For $\xi^j=0$ we have $e^a_i (\xi^0,{\bf 0})=\delta^a_i$.
The fact that the $n$-bein is orthonormal then fixes $e^a_0 (\xi^0,{\bf 0})=
\phi(\xi^0)\delta^a_0$.
The equivalent statement for $\Gamma^a_{~bi}(\xi)$ is reached
by using the same procedure.\\

\noindent Coming back to the proof that the geodesics start orthogonal to the
line $s^\alpha(\xi^0)=x^\alpha(\xi^0,{\bf 0})$, from the previous
result we have that
\begin{equation}
g_{ij}(\xi^0,{\bf 0})=-\delta_{ij}\  \ \ {\rm and}\ \ \
g_{0\mu}(\xi^0,{\bf 0})=\phi(\xi^0)^2 \eta_{0\mu}.
\end{equation}
Thus the following equalities hold
\begin{equation}
\left(\frac{dx}{d\lambda},\frac{ds}{d\xi^0}\right)\biggr |_{\xi^i=0}=
\hat g_{\mu\nu} (x)\xi^i \partial_i x^\mu (\xi^0,\lambda \xi^i)\partial_0
x^\nu(\xi)\biggr |_{\xi^i=0}=g_{i0}(\xi^0,{\bf 0}) \xi^i=0,
\end{equation}
where we have used the rule of transformation of the metric under change of
coordinates. This completes the proof of our statement.

It is interesting that the converse property is true as well. In particular
if we consider the family of geodesics that start orthogonal to the worldline
$s^\alpha (\xi^0)=x^\mu(\xi^0,{\bf 0})$,
they define a function  $x^\mu (\xi^0,\lambda \xi^i)$
that satisfies eqs. (\ref{x(xi)}) and (\ref{Eqq}).
The proof is a straightforward application of the theory of
differential equations.

Another property, which  we are going to use in the following, is that the
FW $n$-bein $e^a(\xi)$ is parallel transported fields along the
geodesics $x^\mu(\lambda)$. In
fact  let us consider the geodesics  $x^\mu(\lambda)$ defined before,
in the FW coordinates they have the simple form $\xi^0={\rm constant}$
and $\xi^i=\lambda v^i$ with $v^i$ constant vector. Then we obtain
\begin{equation}
\frac{D e^a}{d\lambda}=v^i\Gamma^a_{~b i}(\xi(\lambda)) e^b=0,
\end{equation}
where we have used eq. (\ref{form2}). Thus the covariant derivative of
the $n$-bein is zero along this curves. This  means that they are parallel
transported.  This property defines completely the $n$-bein
up to their redefinition  along the line $s^\alpha(\tau)$.

A geometrical construction of this coordinates is now a straightforward
consequence of the previous properties. In fact given a worldline
$s^\alpha(\tau)$, we fix a basis  $E_a(s)$ along the curve with
the following properties\footnote{The fact that $E_0(s)$ must be parallel
to $\frac{ds}{d\tau}$ is a consequence of the previous property.}
\begin{equation}
(E_a(s),E_b(s))=\eta_{ab}\ \ \ \ {\rm  and}\ \ \ \
E_0(s) {\rm \   parallel\  to \ \ }{\frac{ds}{d\tau}}.
\end{equation}
Then the FW coordinates of a generic point with respect to
this curve can be constructed in this way.
Let us consider the geodesic that
starts orthogonal to $s^\alpha(\tau)$ and reaches $P_0$ i.e.
\begin{equation}
x^\alpha(0)=s^\alpha (\tau), \ \ \ x^\alpha(1)=x(P_0 ),\ \ \
(\frac{dx}{d\lambda},\frac{ds}{dt})\vert_{\xi^i=0}=0.
\end{equation}
Using what we have shown before, this geodesics must have the form
\begin{equation}
x^\mu(\lambda)=x^\mu (\xi^0,\lambda\xi^i) ,
\end{equation}
where $x^\mu(\xi)$ is the transformation of coordinates that connects the
coordinates $\{ x^\mu\} $ with the FW coordinates $\{ \xi^\mu\}$.
Then the new coordinates are simply obtained by
\begin{equation}
\xi^0=\tau \ \ \ \ {\rm and}\ \ \ \
\xi^i=
\frac{dx^\mu}{d\lambda}\biggr |_{\lambda=0} E^i_\mu(s(\tau)).
\end{equation}
One can recognize in this construction the usual one. However
 in our case  we have to specify how to construct the
field $e^a_\mu(\xi)$ at each point of the space-time.
The new feature arises from the fact
that we are in
the first order formalism and beyond the diffeomorphisms we have the
local Lorentz invariance. \\
The construction of the FW coordinates associates to each point
$P_0$ a geodesics starting from the wordline $s^\alpha(\tau)$ and  the fields
$e^a_\mu(\xi)$ at the point $P_0$, using  a property previously derived,
are the parallel transported ones along this geodesic of the $n$-bein
$E_a(s)$  that we have at the intersection point of the two curves.
This completes the geometrical construction of the coordinate
system and of the fields.

However there are some  issues that we have to clarify and that are related
to problem of the residual gauge  invariance.  In this construction we make
three arbitrary choices
\begin{itemize}
\item the world-line $s^\alpha(\tau)$ and its parameter $\tau$
\item the affine parameter $\lambda$ for our geodesics $x(\lambda)$
\item the choice of the basis $E_a (s(\tau))$ along the line $s^\alpha(\tau)$
\end{itemize}
The invariance  related to this three points is discussed in \cite{Sem}.
Here we want to make only a few remarks.\\
Eqs. (\ref{form1}) and (\ref{form2}) do not contain any information about
the observer, and in particular its worldline; they express general geometrical
 properties of the FW system which are valid whatever the observer is.
Any information about the observer is an extra degree of freedom that we are
free to fix. From a technical point of view, we can say that we have to
specify  the initial condition
\begin{equation}
x^\mu(\xi,{\bf 0})=s^\mu(\tau)\bigr |_{\tau=\xi^0}
\end{equation}
if we want to solve eqs. (\ref{x(xi)}) and (\ref{Eqq}).\\
Then the possibility of choosing a different affine parameter $\lambda^\prime=
a~\lambda+b$ for our geodesics is only a matter of convenience. It gives
a global rescaling of our coordinates. The choice $x^\mu(0)=s^\mu (\tau)$ and
$x^\mu(1)=x^\mu(P)$ is the usual one.\\
Having a different basis  $E_{a}(s(\tau))$ along the observer' s world
line means that
we can choose the angular velocity for our observer. In other words the
transformation
\begin{eqnarray}
&&\xi^0=\bar\xi^0\ \ \ \ {\rm and}\ \ \ \ \xi^i=\Omega^i_j(\bar\xi^0)\bar \xi^j
\nonumber\\
&& e^0=\bar e^0\ \ \ \ {\rm and }\ \ \ \ e^i=\Omega^i_j(\bar\xi^0)\bar e^j
\end{eqnarray}
where $\Omega$ is a euclidean
rotation in $(n-1)$ dimensional space, preserves the FW structure of
our fields.
It is easy to see that this residual
gauge transformation allows to choose the space-space components of
$\Gamma_0 (\xi^0,{\bf 0})$ vanishing\footnote{We have at our disposal
$(n-2)(n-1)/2$ degrees of freedom in
$\dot\Omega$ which can be use to put to zero the space-space components
of $\Gamma_0({\bf 0},t)$. This is obtained by solving by means of the
standard time ordered integral the equation
$\dot\Omega=-\Gamma~\Omega$ in the $(n-1)\times (n-1)$ space components.}.
We have only the spatial rotation and not a Lorentz transformation because we
are obliged to keep $E_0(s)$ parallel to the observer' s worldline.\\

At the beginning of this section we have recalled that the FW field
are given, in the usual approach, by a perturbative expansion around the
observer' s worldline. The situations changes in this first order formulation.
In fact one can express the fields in term of quadratures of the Riemann and
torsion two forms. In particular one finds \cite{MS1} that
\begin{equation}
\label{A1}
\G^a_{bi}(\xi)=\xi^j\int^1_0 R^a_{bji}(\xi^0,\lambda \bfxi)\lambda d\lambda.
\end{equation}
\begin{equation}
\label{A2}
\Gamma^a_{b0}({\xi})=\G^a_{b0}(\xi^0,{\bf 0})+\xi^i\int^1_0 R^a_{bi0}(\xi^0,
\l {\bfxi})
d\l,
\end{equation}
\begin{equation}
\label{A9}
e^a_0({\bf \xi})=\delta^a_0+\xi^i\int^1_0\Gamma^a_{i0}(\xi^0,\l{\bfxi}) d\l+
\xi^j\int^1_0 S^a_{j0}(\xi^0,\lambda\bfxi)d\l.
\end{equation}

\begin{equation}
\label{A10}
e^a_i(\xi)=\delta^a_i+\xi^j\int^1_0\G^a_{ji}(\xi^0,\l {\bfxi})\l d\l
+\xi^j\int^1_0 S^a_{ji}(\xi^0,\lambda\bfxi) \l d\l
\end{equation}
being $R^a_{bji}$ the curvature and $S^a_{i0}$ the torsion. These formulae
contain the additional hypothesis that $\xi^0$ is identified with the proper
time of the observer. (See footnote pag. 5). The arbitrary functions
$\Gamma^a_{~b0}(\xi^0,{\bf 0})$ that appear in the expression are obviously
related to the residual gauge invariance discussed before.
$\Gamma^0_{~b0}(\xi^0,{\bf 0})$  is the observer' s acceleration and
$\Gamma^i_{~j0}(\xi^0,{\bf 0})$ his angular velocity with respect to
gyroscopic directions \cite{HawE}.

%
\section{CONSTRAINT EQUATIONS FOR THE ENERGY MOMENTUM TENSOR}

The above given treatment is valid in any dimension; in $2+1$
dimensions a simplifying feature intervenes because the Riemann
tensor, being a linear function of the Ricci tensor, can be written
directly in terms of the energy-momentum tensor.
\begin{equation}
\label{einstein}
\varepsilon_{abc} R^{ab}= -{2\kappa} T_c,
\end{equation}
\begin{equation}
R^{ab}=-{\kappa}\varepsilon^{abc} T_c=-{\kappa\over 2}
\varepsilon^{abc}~\varepsilon_{\rho\m\n}\tau^{~\rho}_c dx^\m\wedge dx^\n,
\end{equation}
where $\kappa=8\pi G$ and $T_c$ is the energy momentum two form and
$R^{ab}$ is the curvature two form.

Thus also in the time
dependent case eqs.(\ref{A1},\ref{A2},\ref{A9},\ref{A10}) provide a
solution by quadrature of Einstein's
equations. Nevertheless one has to keep in mind that the solving
formulas are true only in the Fermi-Walker gauge, in which the energy
momentum tensor is not an arbitrary function of the coordinates but is
subject to the covariant conservation law and symmetry property,
that are summarized by the equations
\begin{equation}
{\cal D} T^a=0,
\end{equation}
and by
\begin{equation}
\label{symmetry}
\varepsilon_{abc} T^b\wedge e^c =0.
\end{equation}
It will be useful as done in ref.\cite{MS2} to introduce the
cotangent vectors
$\displaystyle{T_{\mu}=\frac{\partial \xi^0}{\partial \xi^
\mu}}$,
$\displaystyle{P_{\mu}=\frac{\partial \rho}{\partial \xi^
\mu}}$ and
$\displaystyle{\Theta_{\mu}=\rho\frac{\partial \theta}{\partial \xi^
\mu}}$ where $\rho$ and  $\theta$ are the polar variables in the
$(\xi^1,\xi^2)$ plane. In addition  we notice that in $2+1$ dimensions
the most general form of a connection satisfying eq.(\ref{form1}) is
\begin{equation}
\label{connection}
\Gamma^{ab}_\mu(\xi)=\varepsilon^{abc}\varepsilon_{\mu\rho\nu}P^\rho
A^\nu_c(\xi).
\end{equation}
where in eq.(\ref{connection}) obviously the component of $A$ along
$P^\rho$ is irrelevant.

Writing $A^\nu_c$ in the form \cite{MS2}
\begin{equation}
\label{A}
A^\rho_c(\xi)= T_c \left [ \Theta^\rho\beta_1+T^\rho\frac{(\beta_2-1)}{\rho}
\right ]+\Theta_c \left [\Theta^\rho\alpha_1 +T^\rho\frac{\alpha_2}{\rho}
\right ]+P_c \left [\Theta^\rho \gamma_1+T^\rho\frac{\gamma_2}{\rho}
\right ],
\end{equation}
we have for the components $\tau^\rho_c$ of the energy momentum two
form
\begin{eqnarray}
\label{taumisto}
&&\tau^{\rho}_c=-\frac{1}{\kappa}\biggl \{ T_c \biggl ( T^\rho
\frac{\beta^\prime_2}{\rho}+\T^\rho \beta^\prime_1\biggr )+
\T_c
\biggl ( T^\rho
\frac{\alpha^\prime_2}{\rho}+\T^\rho\alpha^\prime_1
\biggr )+
 P_c\biggl (T^\rho \frac{\gamma^\prime_2}{\rho}+
\T^\rho
\gamma^\prime_1\biggr ) +\\
&&\frac{1}{\rho}P^\rho
\biggl [ T_c\biggl (\alpha_1\gamma_2-\alpha_2\gamma_1-
\frac{\dev\beta_1}{\dev \theta}+ {\partial \beta_2\over \partial t}\biggr )+
\T_c \biggl (\beta_1\gamma_2-\beta_2\gamma_1-
\frac{\dev\alpha_1}{\dev \theta}+{\partial \alpha_2\over \partial t}\biggr )
+\\
&&P_c \biggl (\alpha_1\beta_2
-\alpha_2\beta_1-
\frac{\dev\gamma_1}{\dev \theta}+ {\partial \gamma_2\over \partial t}
\biggr )\biggr ]\biggr \}.\nonumber
\end{eqnarray}
The dreibein and the metric are now given by
\begin{equation}
\label{dreibein}
e^a_\mu(\xi)= -T^a (T_\mu A_1+\frac{1}{\rho} \Theta_\mu A_2)-
\Theta^a (T_\mu B_1+\frac{1}{\rho} \Theta_\mu B_2)-P^a P_\mu
\end{equation}
and
\begin{equation}
ds^2= (A_1^2-B_1^2) dt^2 + 2 (A_1 A_2-B_1 B_2)dtd\theta +
(A_2^2-B^2_2) d\theta^2-d\rho^2
\end{equation}
where $A_i$ and $B_i$ are defined by
\begin{eqnarray}
&&A_1(\xi)=\rho \int^1_0 \alpha_1(\lambda \bfxi,t) d\lambda-1\ \ {\  ,\  }
\ \ \ B_1(\xi)=\rho \int^1_0 \beta_1(\lambda \bfxi,t) d\lambda \nonumber,   \\
&&A_2(\xi)=\rho \int^1_0 \alpha_2(\lambda \bfxi,t) d\lambda\phantom{-1}
\ \ {\rm and}
\ \ \ B_2(\xi)=\rho \int^1_0 \beta_2(\lambda \bfxi,t) d\lambda.
\end{eqnarray}
The above equation (\ref{taumisto}) is obtained by substituting
eq.(\ref{A})
into eq.(\ref{einstein}) and as
such the resulting energy-momentum tensor is covariantly conserved.
On the other hand the imposition of the symmetry constraint
eq.(\ref{symmetry}) gives
\begin{mathletters}
\begin{eqnarray}
\label{22a}
&&A_1\alpha_2^\prime -A_2 \alpha_1^\prime
+B_2\beta_1^\prime
 -B_1 \beta_2^\prime=0\\
\label{22b}
&&\alpha_2 \gamma_1-\alpha_1 \gamma_2+A_2\gamma_1^\prime -A_1
\gamma_2^\prime
+\frac{\partial\beta_1}{\partial\theta}-{\partial\beta_2\over \partial t}=0\\
\label{22c}
&&\beta_2 \gamma_1-\beta_1 \gamma_2+B_2\gamma_1^\prime -B_1
\gamma_2^\prime +\frac{\partial\alpha_1}{\partial\theta}-
{\partial\alpha_2\over \partial t}=0.
\end{eqnarray}
\end{mathletters}
These relations can be integrated with respect to $\rho$ and taking into
account the regularity conditions of $\alpha_i$, $\beta_i$ and $\gamma_i$
at the origin we reach
\begin{mathletters}
\label{symmetryeq}
\begin{eqnarray}
\label{symmetry1}
&&A_1\alpha_2-A_2 \alpha_1+B_2
\beta_1-B_1\beta_2=0\\
\label{symmetry2}
&&A_2\gamma_1 -A_1
\gamma_2 +\frac{\partial B_1}{\partial\theta}-
{\partial B_2\over \partial t}=0\\
\label{symmetry3}
&&B_2\gamma_1-B_1
\gamma_2 +\frac{\partial A_1}{\partial\theta}-{\partial A_2\over
\partial t}=0.
\end{eqnarray}
\end{mathletters}
\section{Time dependent case with axial symmetry}
In presence of axial symmetry eq.(28) simplify to

\begin{mathletters}
\begin{eqnarray}
\label{23a}
&&A_1\alpha_2^\prime -A_2 \alpha_1^\prime
+B_2\beta_1^\prime
 -B_1 \beta_2^\prime=0\\
\label{23b}
&&\alpha_2 \gamma_1-\alpha_1 \gamma_2+A_2\gamma_1^\prime -A_1
\gamma_2^\prime-{\partial\beta_2\over \partial t}=0\\
\label{23c}
&&\beta_2 \gamma_1-\beta_1 \gamma_2+B_2\gamma_1^\prime -B_1
\gamma_2^\prime-{\partial\alpha_2\over \partial t}=0.
\end{eqnarray}
\end{mathletters}

Eq.(\ref{taumisto}) implies that
outside the sources the following quantities are constant both with
respect to $\rho$ and $t$
\begin{equation}
\label{qua2}
\beta_2^2-\alpha_2^2-\gamma_2^2= {\rm const}.
\end{equation}

\begin{equation}
\label{qua3}
\alpha_2 B_2-\beta_2 A_2= {\rm const}.
\end{equation}
We shall relate such conserved quantities with the energy
and the angular momentum of the system.

The procedure for proving the independence both on $\rho$ and $t$ of
eq.(\ref{qua2}) and eq.(\ref{qua3}) is completely similar to the one
used to prove the
constancy in $\rho$ and $\theta$ of $\beta_1^2-\alpha_1^2-\gamma_1^2$
and $\alpha_1 B_1-\beta_1 A_1$ in the stationary case for which we refer to
\cite{MS2}.

We come now to the computation of the Lorentz holonomy of the connection
(\ref{connection}) along a circle at constant time $t$ and which
encompasses all matter.
It is of the form $W_L=e^{iJ_a\Delta^a}$
where $J_a$ are the generators of the Lorentz group
in the fundamental representation with commutation relations
$[J_a,J_b] = i \varepsilon_{ab}^{~~c} J_c$, and it is obtained by
substituting eq.(\ref{connection}) and (\ref{A}) into
\begin{equation}
{\rm Pexp}\left [-i \oint J_a \Gamma^a_i ~dx^i\right ]
 \equiv e^{iJ_a \Delta^a}.
\end{equation}
We have
\begin{equation}
J_a\Gamma^a_1 dx^1 + J_a\Gamma^a_2 dx^2 = [-J_0 (1-\beta_2) - (J_2
\cos\theta-J_1 \sin\theta)\alpha_2 - (J_1 \cos\theta + J_2
\sin\theta)\gamma_2] d\theta
\end{equation}
from which using ref.\cite{VolDol} we have
\begin{equation}
{\rm Pexp}(-i \int J_a \Gamma^a_i dx^i) = \exp(i J_0\theta)
{\rm Pexp}(-i \int J_a V^a d\theta)
\end{equation}
with $V^a=(\beta_2,-\gamma_2,-\alpha_2)$ which for $\theta=2\pi$ and $V$
time-like becomes
\begin{equation}
\exp(2\pi i J_a\hat V^a (1-V))
\end{equation}
with $V=\sqrt{\beta_2^2-\alpha_2^2-\gamma_2^2}$. Then the angular
deficit and the mass are given by
$2\pi\left (1-\sqrt{\beta_2^2-\alpha_2^2-\gamma_2^2}\right )= 8 \pi G M$.

Coming now to the Poincar\'e holonomy we have, with $P_a$ the
translation generators of $ISO(2,1)$ in the fundamental representation
\begin{eqnarray}
&& W_P=e^{iJ_a \Delta^a + i P_a \Xi^a}=
e^{2\pi i J_a \hat V^a } e^{-2\pi i(J_0\beta_2 -J_1\gamma_2 - J_2\alpha_2 +P_2
B_2-P_0 A_2)} = \\
&& e^{2\pi i J_a \hat V^a} e^{-2\pi i
(J_0 \beta_2 -J_1 \gamma_2 - J_2\alpha_2)}
e^{-2\pi i(P_2
\tilde B_2-P_0 \tilde A_2)}
\end{eqnarray}
with $\tilde B_2 \alpha_2 - \tilde A_2 \beta_2 = B_2 \alpha_2 - A_2
\beta_2 $.
Such a quantity, as discussed in \cite{MS3} \cite{AcTo} \cite{Witt},
is an invariant under gauge transformations and in particular  under
deformations of the loop which do not intersect matter. Then we can
rewrite
the holonomy in the form
\begin{equation}
W_P= \exp( 2\pi i J_a\hat V^a (1-V)) \exp(-2\pi i(P_2
\tilde B_2-P_0 \tilde A_2))
\end{equation}
and the angular momentum is related to the written invariant according
to the general formula \cite{MS3}
\begin{equation}
{\cal J} = {\Delta_a\Xi^a \over 8\pi G \sqrt{\Delta_a\Delta^a}}=
{\alpha_2 B_2 -\beta_2 A_2\over 4 G V}.
\end{equation}
Choosing $\alpha_2,\beta_2,\gamma_2$ satisfying
eq.(\ref{qua2},\ref{qua3}) as input functions specifying an
energy-momentum tensor with bounded support it is possible to write
down using a procedure completely similar to that used in the
stationary case \cite{MS2}, quadrature formulas which express the
metric in term of the $\alpha_2,\beta_2,\gamma_2$.

\section{STATIONARY CASE}
The stationary problem in the Fermi-Walker gauge has been introduced
in \cite{MS1}\cite{MS2}. A major difficulty of the applicability of
the Fermi-Walker
coordinates is whether they give a complete description of the space
time manifold. For the time dependent case we know that in general
this is not the case \cite{MiThWe}. On the other hand in the
stationary case there exists a powerful projection technique due to
Geroch that allows to infer from the completeness of such a
projection the completeness of the Fermi-Walker coordinate system.

Given a Killing vector field, that in our case will be assumed to be
everywhere time-like and never vanishing, one defines the Geroch
projection $S$ \cite{geroch} of the $2+1$ manifold $M$ as the quotient
space of $M$ by the motion generated by the Killing vector field; in
different words $S$ is the collection of all trajectories in $M$ which
are everywhere tangent to the Killing vector field. We shall assume
following \cite{geroch} that $S$ has the structure of a differentiable
manifold; then Geroch \cite{geroch} shows that $S$ is endowed of a
metric structure with metric tensor given by
\begin{equation}
h_{ab}=g_{ab}- {K_a K_b\over K^2}
\end{equation}
which, being the Killing vector $K$ time-like, is negative definite
(space-like metric).


If such a manifold is metrically complete, (if it is not we can
consider its metric completion)  then we know from the
theorem of Hopf-Rinew-De Rahm \cite{spivak}, that it is also geodesically
complete. This means that given two points is $S$ there exists always
at least one geodesic connecting them. Given an event in $M$ we shall
consider the geodesic in $S$ which connects its projection with the
projection of the world line of the stationary observer.
We want now to relate such geodesics on the Geroch projection $S$ to
the geodesics in the $2+1$ dimensional manifold $M$, that where used
to define the Fermi-Walker coordinate system. To this end let us
consider the integral curve starting from our event, of the vector
field on $M$ which is orthogonal to the Killing vector $K$ and
possesses as Geroch projection the tangent vector to the geodesic in
$S$. Then we have
\begin{equation}
v^\alpha K_\alpha=0,~~~~~~~~v^\mu h^\alpha_{\beta}\nabla_\mu v^\beta=0.
\end{equation}
But then, keeping in mind that
\begin{equation}
h^\alpha_\beta=\delta^\alpha_\beta - K^\alpha K_\beta/K^2
\end{equation}
and that
\begin{equation}
v^\mu K_\alpha\nabla_\mu v^\alpha=-v^\mu v^\alpha \nabla_\mu K_\alpha=
v^\mu v^\alpha \nabla_\alpha K^\mu=0
\end{equation}
we have $v^\mu\nabla_\mu v^\alpha=0$ i.e. the considered curve is a
geodesic in
$M$. Such geodesic will meet the line $O$ at a certain  time $t$ and
will be orthogonal to it.
Thus we have reached the conclusion that the Fermi-Walker coordinate
system constructed in \cite{MS1} is complete, and possibly
overcomplete.

In ref.\cite{MS2} we proved algebraically, starting from eq.(28) and
eq.(\ref{taumisto})  for the general stationary problem, that
outside the sources we have two invariants i.e. the expressions
\begin{equation}
\beta_1^2-\alpha_1^2-\gamma_1^2 = C_1
\end{equation}
and
\begin{equation}
\alpha_1 B_1 - \beta_1 A_1 = C_2
\end{equation}
become independent of $\rho$ and $\theta$, outside the sources. We
shall give here $C_1$ and $C_2$ an interpretation in term of Lorentz
and Poincar\'e holonomies.

Let us consider a Wilson loop that has two branches $AB$ and $CD$
parallel to the time Killing vector field  and of unit length in the
Killing time and connect them by two
arcs $BC$ and $DA$ each of which develops at constant time. We have
\begin{equation}
W= W_{AD}W_{DC}W_{CB}W_{BA}=1
\end{equation}
if the whole Wilson loop is taken outside the sources. Then as the two
Wilson arcs $W_{CB}$ and $W_{DA}$, owing to the stationary nature of the
problem are equal $W_{CB}=W_{DA}=U$, we have
\begin{equation}
W_{BA}= U^{-1} W_{CD} U.
\end{equation}
In words, two Wilson lines parallel to time and translated in $\rho$
and $\theta$ without intersecting the sources are related by a
similitude transformation. Let us compute now $W_{BA}$.
We have
\begin{equation}
J_a\Gamma^a_0= J_0 \beta_1-\tilde J_1\gamma_1 -\tilde J_2\alpha_1
\end{equation}
and
\begin{equation}
P_a e^a_0= -P_0 A_1+\tilde P_2 B_2
\end{equation}
where $\tilde J_1= \cos\theta J_1+ \sin \theta J_2$, $\tilde J_2=
\cos\theta J_2 - \sin \theta J_1$, $\tilde P_1= \cos\theta P_1+
\sin \theta P_2$  and $\tilde P_2= \cos\theta P_2 - \sin \theta P_1$.
As $J_0, \tilde J_1  , \tilde J_2 , P_0, \tilde P_1  , \tilde P_2$
satisfy the same commutation relations as  $J_0, J_1, J_2, P_0, P_1,
P_2$ we have the
two invariants $C_1$ and $C_2$ given by the combinations
\begin{equation}
\Delta^a \Delta_a=2\pi (\beta_1^2-\alpha_1^2-\gamma_1^2)~~~~~~~~
{\rm and}~~~~~~~~ \Delta^a \Xi_a=2\pi (A_1\beta_1-B_1\alpha_1).
\end{equation}
In appendix A the two written invariants are expressed in terms the norm
of the vorticity of the Killing vector and the projection of the curl
of the Killing vector along the Killing vector itself.

%
%
\section{The problem of closed time-like curves in the stationary
case}

The problem of CTC's in the stationary case for open universes in
presence of  axial
symmetry has been dealt with in ref.\cite{MS2}.  The result is
simple to state:
If the matter sources satisfy the weak energy condition (WEC) , the
universe in open
and there are no CTC at space infinity, in presence of axial symmetry
there are no CTC
at all. We recall in addition that explicit examples show that the
hypothesis of absence of CTC's at space infinity is a necessary one.

Here we shall give a simplified treatment that extends the result to
closed universes
always with axial symmetry, and that  under a certain assumption about
the behavior
of the determinant of the dreibein in the Fermi-Walker gauge, extends also to
stationary universes in absence of axial symmetry.

First we notice that if for $\rho >0$, $A_2^2-B_2^2\equiv
g_{\theta\theta} <0$ there
cannot be CTC's.
In fact given the closed curve $\xi(\lambda)$ let us consider a point
where $\displaystyle{{\partial \xi^0\over \partial\lambda}=0}$. There we have
\begin{equation}
ds^2=(A_2^2-B_2^2) d\theta^2-d\rho^2
\end{equation}
and if $ A_2^2-B_2^2 \leq 0$ it cannot be an element of a CTC.
Then
also for the non axially symmetric stationary problem, it is enough
to prove that for
$\rho >0$, $ A_2^2-B_2^2 \leq 0$. Without committing ourselves to the axially
symmetric case we shall start proving the following lemma.

Lemma: If the WEC holds, $\det(e)>0$ for $\rho>0$ and $g_{00}\equiv
A_1^2-B_1^2 >0$ (and
thus never vanishes), and the space is conical at infinity then
there are no CTC at all.

We notice that $g_{00}\equiv A_1^2-B_1^2 >0$ express the requirement of the
existence of a non singular time-like Killing vector field.

Proof: $ A_2^2-B_2^2$ can vanish either for $A_2 = B_2$  (zero of type
$+$)  or
for $A_2 = -B_2$  (zero of type $-$). If $ A_2^2-B_2^2$, which is zero at the
origin and due to the behavior of $A_2$ and $B_2$ is negative in a
neighborhood of the  origin, changes sign for a certain $\theta$ and
$\rho$ it has to revert, as $\rho$ increases, to the negative
sign according to the results of Appendix C.
Let us consider the first zero
of $A_2^2- B_2^2$ after which it becomes again negative and suppose
this zero to be of type $+$,
i.e. $A_2=B_2$ and let call it $\rho_+$. Then in $\rho_+$ we must have for
$g_{\theta\theta}$ a non positive derivative i.e.
\begin{equation}
A_2(\alpha_2-\beta_2)\leq 0
\end{equation}
with $A_2\neq 0$ because we cannot have  $A_2=B_2=0$ otherwise $\det(e)=0$.
We also have
\begin{equation}
\det(e)=A_2(B_1-A_1)>0
\end{equation}
which gives as a consequence
\begin{equation}
(\alpha_2-\beta_2)(B_1-A_1)\leq 0.
\end{equation}
Then defined
\begin{equation}
\label{WECdelta}
E^{(\pm)} (\rho)\equiv
(B_2\pm A_2)(\alpha_1\pm\beta_1)-
(\alpha_2\pm\beta_2)(B_1\pm A_1)
\end{equation}
in $\rho_+$ we have
\begin{equation}
E^{(-)}(\rho_+)= -(\alpha_2-\beta_2)(B_1-A_1)\geq 0.
\end{equation}
We recall however \cite{MS2} that as a direct consequence of the WEC,
$E^{(\pm)}$ are non increasing functions of $\rho$. This fact implies
$E^{(-)}(\rho <\rho_+)\geq 0$.
Now we consider the following identity
\begin{equation}
{d\over d\rho}({B_2-A_2\over B_1-A_1}) = { E^{(-)}(\rho)\over (B_1-A_1)^2}
\geq 0
\end{equation}
for $\rho <\rho_+$. We recall in addition that $ B_1-A_1>0$ because it cannot
vanish and at the origin equals $1$. But this implies that $A_2-B_2$
is identically
$0$ from the origin to $\rho_+$ which contradicts the fact that in a
neighborhood of
the origin the same quantity has to be negative. Then the above described zero
at $\rho_+$ cannot exist. Similarly one reasons for a zero of type $-$
and we reach the
conclusion that $A_2^2-B_2^2$ has to be always negative except at the
origin where has the value $0$.

This lemma is already sufficient to exclude CTC for open universes whenever
$\det(e)$ never vanishes for $\rho>0$. The non vanishing of $\det(e)$
for $\rho >0$ is a rather strong requirement in
absence of axial symmetry. On the other hand it was proved in
ref.\cite{MS2}  that in
presence of axial symmetry the vanishing of $\det(e)$ leads either to
the closure of the universe or to the compactification of the
$2+1$ dimensional manifold (see Appendix D).
Referring now to the case of axial symmetry we extend the result on the
absence of CTC's to closed universes. We shall prove in what
follows that the
WEC plus axial symmetry implies the absence of CTC's in any closed
stationary universe.

If the universe closes ( with the topology of a sphere due to the
axial symmetry ) then for a certain
$\rho_0$ we must have $\det(e)(\rho_0)=0$ as it is imposed by the
vanishing of the component
$\gamma_{\theta\theta}$ of the space metric and in addition (see
Appendix D)  in $\rho_0$ $A_2^2-B_2^2=0$.
In $\rho_0$ we
must have necessarily $A_2=B_2=0$
otherwise substituting into $\det(e)=0$ we would get either $B_1-A_1=0$ or
$B_1+A_1=0$ which would make the time-like Killing vector field
singular at that point.
We notice furthermore that in $\rho_0$, $(A_1^2-B_1^2)
(\alpha_2^2-\beta_2^2) <0$ as can be
seen form the symmetry equation (\ref{symmetry1}) at $\rho_0$ written
in the form
\begin{equation}
(\alpha_2+\beta_2)(A_1-B_1)=-(\alpha_2-\beta_2)(A_1+B_1)
\end{equation}
and being $A_1^2-B_1^2>0$ we have $\alpha_2^2-\beta_2^2 <0$. This
means that in the neighborhood of $\rho_0$ there cannot be CTC's and
thus we are under the same hypothesis of the proof for open axially
symmetric universes.

\section{Conclusions}
The Fermi-Walker gauge in 2+1 dimensional gravity has
been successful both in dealing with extended sources and time
dependent problems. In this gauge it is possible to write down general
resolvent formulas that contain only quadratures and express the
metric in term of the source of the gravitational field i.e. the
energy-momentum tensor. When a Killing vector exists (axially symmetric
problem  or stationary problem ) it is also possible to treat
explicitly the support of the energy-momentum tensor. The compactness
condition on the sources can be expressed algebraically in terms of
the Poincar\'e holonomies which in the stationary case can be related
to the vorticity of the Killing vector.
We have proven here that for the stationary problem the completeness
of the Geroch projection implies the completeness of the Fermi-Walker
coordinate system. In this context we gave an extension of the theorem
on the absence of CTC \cite{MS2} to the case of closed universes, with
axial symmetry.  In addition whenever the determinant of the dreibein
in the Fermi-Walker system does not vanish, the proof extends also to
the stationary case in absence of axial symmetry.

\typeout{inizio}
\appendix
\section{}
In this appendix we shall review how eqs. (\ref{x(xi)}) and (\ref{Eqq})
constrain the form of the function $x^\mu(\xi^0,\xi^i)$. Eq. (\ref{x(xi)})
can be transformed in an integral one through a standard procedure. It
becomes
\begin{equation}
\label{lkj}
x^\mu(\xi^0,\xi^i)=s^\mu(\xi^0)+J^\mu_i (\xi^0)\xi^i +
\int^1_0 d\alpha~ (1-\alpha)~ \xi^i \partial_i x^\rho(\xi^0,\alpha\bfxi)
\xi^j\partial_j x^\sigma(\xi^0,\alpha\bfxi)\hat
\Gamma^\mu_{\rho\sigma}(x(\xi^0,
\alpha\bfxi)).
\end{equation}
$s^\mu(\xi^0)$ and $J^\mu_i (\xi^0)$ are the initial values at
$\xi^i=0$.
They correspond to the observer' s trajectory and to the Jacobian of the
transformation along the line. This equation can be solved recursively. The
existence of the solution is assured, at least locally, under the same
assumptions that guarantees the existence of the solution of geodesic
equation. Due to the nature of eq. (\ref{x(xi)}) it would be possible to
consider $s^\mu$ and $J^\mu_i$ that are homogeneous function of degree
zero in the variables $\xi$. This is ambiguity is avoided by looking
for regular solution ($C^2$ or better) of eq. (\ref{lkj}). In fact this other
choice would lead  to solution that are singular at the origin.\\

The next step is to impose eq. (\ref{Eqq}). Actually eq. (\ref{Eqq}) is
not a true differential equation but it can be rewritten as a constraint
on the initial data. Let us define
\begin{equation}
L(\xi)=\xi^j\frac{\partial x^\mu}{\partial \xi^j} \hat g_{\mu\nu}(x(\xi))
\xi^i\frac{\partial x^\nu}{\partial \xi^i}.
\end{equation}
Using eq. (\ref{x(xi)}) it is easy to show that this quantity satisfies
\begin{equation}
\xi^i\frac{\partial L(\xi)}{\partial\xi^i}=2 L(\xi).
\end{equation}
This means  that $L(\xi)$ is a homogeneous function of degree $2$
in the $\xi^i$ variables. Every function of this kind can be written like
$L(\xi)=C_{ij}(\xi)\xi^i\xi^j$ where $C_{ij}(\xi^0,\xi^i)$ is a homogeneous
function of degree $0$ in the $\xi^i$ variables. However the regularity
of the solution is preserved only if  $C_{ij}$ is independent of $\xi^i$.
At the end we have
\begin{equation}
L(\xi)=\xi^j\frac{\partial x^\mu}{\partial \xi^j} \hat g_{\mu\nu}(x(\xi))
\xi^i\frac{\partial x^\nu}{\partial \xi^i}=C_{ij}(\xi^0)\xi^i\xi^j.
\end{equation}
Now from the explicit form of $L(\xi)$ we can show that
\begin{equation}
C_{ij}(\xi^0)=
J_i^\mu(\xi^0) \hat g_{\mu\nu}(x(\xi^0,{\bf 0})) J_j^\nu(\xi^0).
\end{equation}
Then the eq. (\ref{Eqq}) is simply solved if we impose
\begin{equation}
\xi^j J_j^\mu(\xi^0) \hat g_{\mu\nu}(x(\xi^0,{\bf 0}))
\xi^i J_i^\nu(\xi^0)=-\sum_i\xi^i \xi^i.
\end{equation}
As we see this is a constraint on the possible initial condition $J_i^\mu$.

\section{}
In this appendix we shall point out the relation between the two invariants
$C_1$ and $C_2$, which we have found in the stationary case, and the usual
Geroch formalism \cite{geroch}. This is useful to understand
their geometrical meaning. Following ref. \cite{geroch}, we introduce
the vector
\begin{equation}
 \omega_\alpha=\frac{1}{2}\epsilon_{\alpha\beta\gamma}\nabla^\beta K^\gamma,
\end{equation}
where $K^\gamma$ is the Killing vector of our metric and
$\epsilon_{\alpha\beta\gamma}= \sqrt{g}~
\varepsilon_{\alpha\beta\gamma}$. From $\omega_\alpha$ and
the Killing vector we can construct two scalars
\begin{equation}
\omega_1=\omega_\alpha \omega^\alpha ~~~~~~~{\rm and}~~~~~~~ \omega_2=
K_\alpha\omega^\alpha.
\end{equation}
Using the well-known relation $\nabla_\alpha\nabla_\beta K_\gamma =
R_{\delta \alpha,\beta \gamma} K^\delta$
and the fact that $R_{\alpha\beta,\gamma\delta}=0$ outside the source,
it is straightforward
to show
\begin{equation}
\nabla_\alpha \omega_1=0~~~~~~~{\rm and}~~~~~~~ \nabla_\alpha \omega_2=0,
\end{equation}
i.e. $\omega_1$ and $\omega_2$ are constant outside the source. \\
Now if we express $\omega_1$ in our reference frame we obtain
\begin{equation}
\label{ooo}
\omega_1={1\over 2}\nabla_\alpha K_\beta\nabla^\alpha K^\beta=
{1\over 2}\Gamma^a_{b0}\Gamma^b_{a0}=
\beta_1^2-\alpha_1^2-\gamma_1^2=C_1.
\end{equation}
In the case of $\omega_2$ we have
\begin{equation}
\omega_2=\frac{1}{2}\epsilon_{\alpha\beta\gamma}K^\alpha \nabla^\beta
K^\gamma = {{\rm det}(e)}\epsilon_{0\mu\lambda}g^{\mu\sigma}
\Gamma^\lambda_{\sigma 0}=A_1 \beta_1-B_1 \alpha_1=C_2.
\end{equation}
Then $C_1$ is a constraint on the gradient of $X=K^\alpha K_\alpha$;
in fact eq.
(\ref{ooo}) can be rewritten in the following way
\begin{equation}
\nabla_\alpha X \nabla^\alpha X = 4 (C_1 X -(C_2)^2).
\end{equation}
The meaning of $C_2$ is more clear.  $C_2=0$ corresponds to the
condition of local integrability of our Killing vector. Thus
if $C_2$ is different from zero $K^\alpha$ is not locally integrable.
Finally
we notice that $C_2=0$ does not assure that the problem is static, in fact
in order to have a static problem the Killing vector must be globally
integrable, that is there must exist a family of surfaces orthogonal to
$K^\alpha$. (E.g. the ordinary ``Kerr'' solution in 2+1 dimensions
has $C_2=0$, but we cannot construct a family of surfaces orthogonal
to the Killing vector).

In the stationary case it is possible to formulate
the CTC problem in terms of invariant quantities related to $K$, as
its norm $K^2$ and the vorticity $\omega_2$.

Given a closed curve $C$ in $2+1$ dimensions we can consider the
invariant time-shift defined by
\begin{equation}
\Delta = \oint {K\cdot dx\over K^2}
\end{equation}
which in a one valued coordinate system can also be rewritten as
\begin{equation}
\Delta = \oint (dt +{g_{0i}\over g_{00}}dx^i)=
\oint {g_{0i}\over g_{00}}dx^i.
\end{equation}
Consider now the Geroch (bidimensional) projection $\bar C$ of the
curve $C$.
The necessary and sufficient condition for the
existence of a curve which is a CTC and has as Geroch projection $\bar
C$ is
\begin{equation}
\label{equa}
\Delta \equiv \oint {K\cdot dx\over K^2} > \oint {dl\over K}
\end{equation}
where $dl$ is the length of the line element of $\bar C$ given by
\begin{equation}
dl = \sqrt{\gamma_{ij} dx^i dx^j} =\sqrt{\left  ({g_{0i}g_{0j}\over
g_{00}} - g_{ij} \right ) dx^i dx^j}.
\end{equation}
In fact if $(dt,dx^i)$ is time-like we have
\begin{equation}
g_{00}(dt+ {g_{0i}\over g_{00}} dx^i)^2 - \gamma_{ij} dx^i dx^j >0
\end{equation}
which for $K^2\equiv g_{00} >0$, as we have, gives
\begin{equation}
dt + {g_{0i}\over g_{00}} dx^i > \sqrt{{ \gamma_{ij} dx^i dx^j\over
g_{00}}}= { dl\over \sqrt{g_{00}}}
\end{equation}
and thus eq.(\ref{equa}).

Viceversa suppose $C$ is a closed curve for which eq.(\ref{equa}) is
satisfied. Then given its Geroch projection let us consider the lifting
of $\bar C$ to a future directed light-like curve i.e. with
\begin{equation}
dt+ {g_{0i}\over g_{00}} dx^i= {dl \over \sqrt{g_{00}}}.
\end{equation}
If eq.(\ref{equa}) is satisfied then we have
\begin{equation}
\oint dt = \oint {dl\over K} - \oint {K\cdot dx \over K^2} <0
\end{equation}
which implies the existence of a CTC. It is interesting
that the l.h.s. of eq.(\ref{equa}) can be written in terms of
the vorticity of the Killing vector.

In fact by using the defining property of the Killing vector field
$\nabla_a K_b + \nabla_b K_a =0$ one easily proves that the dual of
the curl of
the field $K^a/K^2$ is parallel to $K^a$ itself i.e.
\begin{equation}
\nabla_a ({K_b\over K^2}) - \nabla_b ({K_a\over K^2})
= 2 \epsilon_{abc} K^c {\omega_2 \over (K^2)^2}
\end{equation}
and thus
\begin{equation}
\oint {K\cdot dx \over K^2}= \int_\Sigma \sqrt{g}~ \epsilon_{abc} {K^c
\omega_2 \over (K^2)^2}   dx^a \wedge dx^b =
\int_\Sigma \sqrt{\gamma}~ \epsilon_{abc} { K^c
\omega_2 \over (K^2)^{3/2}}  dx^a \wedge dx^b=
\end{equation}
\begin{equation}
= 2 \int_\Sigma {\omega_2 \over (K^2)^{3/2}} d\Sigma
\end{equation}
being $d\Sigma$ the area element of the Geroch projection. In
conclusion one can state the necessary and sufficient condition for
the existence of a CTC as
\begin{equation}
\oint {dl\over K} < 2 \int_\Sigma
{\omega_2 \over (K^2)^{3/2}} d\Sigma.
\end{equation}

\section{}
In this appendix we shall show some relevant properties of the metric
\begin{eqnarray}
\label{mtr}
ds^2=&&(A_1(\rho,\theta)^2-B_1(\rho,\theta)^2) dt^2+2 (A_1(\rho,\theta)
A_2(\rho,\theta)-B_1(\rho,\theta)B_2(\rho,\theta))d\theta dt-\nonumber\\
&&d\rho^2+(A_2(\rho,\theta)^2-B_2(\rho,\theta)^2)d\theta^2
\end{eqnarray}
under the following two hypothesis: I) $g_{00}(\rho,\theta)$ is positive
at space infinity, i.e. $t$ is a good time far from the source; II)
the metric (\ref{mtr})  is conical at space infinity, i.e. we can find,
far from the source, a reference frame $\{\tau,r,\phi\}$
where it assumes the reduced form
\begin{equation}
ds^2=(d\tau+J d\phi)^2-d r^2-\alpha^2 r^2 d\phi^2.
\end{equation}
In the following we shall work outside the source.

\noindent
An immediate consequence of II is that the two invariants $C_1$ and $C_2$
defined in sect. IV are zero. Instead the hypothesis I imposes that
$(\alpha^0_1)^2-(\beta^0_1)^2\ge 0$ outside  the source.
Combining this inequality with $C_1=0$ we obtain
\begin{equation}
\alpha^0_1(\theta)=\pm\beta^0_1(\theta)~~~~~~{\rm and}~~~~~~
{\gamma^0_1(\theta)=0}
\end{equation}
Now for $\alpha^0_1=\pm\beta^0_1\not =0$, using $C_2=0$ we get
$A^0_1(\theta)=\pm B^0_1(\theta)$. This result with the previous one
imposes that $g_{00}$ is always zero for all $\rho$ outside the  source,
which contradicts hypothesis I. Thus we are forced to put
$\alpha^0_1=\beta^0_1=0$. With this choice the metric assumes the
simplified form
\begin{equation}
\label{simp1}
ds^2 = g_{00}(\theta) dt^2 + 2g_{0\theta}(\rho,\theta)dt
d\theta-g_{\theta\theta}(\rho,\theta)d \theta^2 -d\rho^2.
\end{equation}
Further simplifications can be obtained, taking into account
symmetry equations. Specifically, we shall show
that $g_{0\theta}$ depends only on $\theta$ and not on $\rho$
and that $g_{00}$ is a constant independent of $\theta$. We have
\begin{eqnarray}
&&g_{0\theta}=A_1A_2-B_1B_2 =(\rho-\rho_0)^2(\alpha_1\alpha_2-\beta_1\beta_2)+
\nonumber \\
&&(\rho-\rho_0)(\alpha_1 A^0_2+\alpha_2 A^0_1-\beta_1 B^0_2-\beta_2
B^0_1)+(A^0_1A^0_2 -B^0_1B^0_2)=(A^0_1A^0_2 -B^0_1B^0_2)
\end{eqnarray}
where the linear term vanishes owing to the eq. (\ref{symmetry1}).
The other two symmetry equations (\ref{symmetry2}) and (\ref{symmetry3})
imply that
\begin{equation}
\frac{\partial g_{00}}{\partial \theta}=
{\partial (A_1^2-B_1^2)\over \partial\theta}=2~{\rm det}(e)\gamma_1=0
\end{equation}
where we have used the fact that $\gamma_1(\theta)=0$.

\section{}

To make the treatment of the CTC in this paper self contained, we
summarize here some basic results which are found in \cite{MS2}.
\begin{equation}
{d E^{(\pm)}(\rho)\over d\rho} \leq 0
\end{equation}
is given by the WEC computed on the two light-like vectors $T^a\pm
\Theta^a$ using the expression of the energy-momentum tensor in the
internal space ${\cal T}_{ab}=\tau^\rho_a e_{b\rho}$ as given by
eq. (\ref{taumisto}) combined with eq. (\ref{dreibein}).

With regard to $\det(e)$, the trace of the energy-momentum tensor is
given by
\begin{equation}
\label{mt}
T^\mu_\mu= -{1\over \kappa}\left [{(\det (e))''\over \det (e)}+
{\alpha_1\beta_2 -\alpha_2\beta_1 \over \det (e)}\right ]=
-{1\over \kappa}\left [{(\det (e))''\over \det (e)}+ \lambda_2\right ]
\end{equation}
being $\lambda_2$ the last eigenvalue of the energy-momentum
tensor. The regularity of the two scalars $T^\mu_\mu$ and $\lambda_2$
imply that if $\det(e)$ vanishes in $\rho_0$ then in the neighborhood
of $\rho_0$ we have
\begin{equation}
\det(e)= cr[1+O(r^2)],
\end{equation}
where $r=(\rho_0-\rho)$.

If in $\rho_0$ $A_2^2 - B_2^2 =0$ then the vanishing of the
determinant imposes $A_1A_2 -B_1 B_2 =0$ which combined with the fact
that the norm of the Killing vector $A_1^2 - B_1^2$ by hypothesis is
always positive, gives $A_2^2 = B_2^2 = 0 $ in $\rho_0$  which fed into
the symmetry equation eq.(\ref{symmetry1}) gives
$A_1 \alpha_2 - B_1 \beta_2 =0$.
Then the fact that for $0<\rho<\rho_0$~ $\det e >0$, tells us, due to
$A_1 \alpha_2 - B_1 \beta_2 =0$, that $\alpha_2^2 - \beta_2^2 <0$.
But for $\alpha_2^2-\beta_2^2<0$
the universe spatially closes with the topology of a sphere;
to avoid a cusp singularity at $\rho_0$ i.e. to have a regular closure
we must have $\alpha_2^2-\beta_2^2 = -1$.

If on the other hand in $\rho_0$ we have $A_2^2-B_2^2 \neq 0$ and due
to $A_1^2 - B_1^2 >0$ necessarily $A_2^2-B_2^2 > 0$; then by means of a
rotation with constant angular velocity, we reduce the metric
around the point $\rho_0$ to the form
\begin{equation}
ds^2= r^2 (\alpha_1^2-\beta_1^2)  dt^2+2 r^2
(\alpha_1\alpha_2-\beta_1\beta_2) d\theta dt+(A^2_2-B^2_2)d\theta^2
-dr^2.
\end{equation}
with  $(\alpha_1^2-\beta_1^2)(A_2^2-B_2^2) <0$ (strict inequality) due
to eq. (\ref{symmetry1}) and thus
$\alpha_1^2(\rho_0)-\beta_1^2(\rho_0) <0 $.
The transformation which regularizes the metric is
\begin{equation}
x=r\cos\sqrt{\beta_1^2-\alpha_1^2}~t,\ \ \ \
y=r\sin\sqrt{\beta_1^2-\alpha_1^2}~t
\end{equation}
and thus $t$ becomes a compact variable, $r$ is restricted to $r>0$
i.e. $0<\rho<\rho_0$ and the universe becomes a compact three
dimensional manifold.


\end{document}